\documentstyle{elsart}

\begin{document}
\begin{frontmatter}
\title{Cosmic ray composition estimation below the knee of the spectrum
from the time structure of \v{C}erenkov light in EAS}
\author{M.D.~Roberts}
\address{Institute for Cosmic Ray Research, University of Tokyo,
Tokyo~188-8502, Japan}

\begin{abstract}
Monte Carlo simulations show that the pulse profile of \v{C}erenkov photons
measured near the core of an extensive air shower is sensitive to the 
secondary muon/electron ratio of the cascade. 
\v{C}erenkov pulses can easily be measured with a single large area
mirror viewed by a photomultiplier tube subtending a small field of view
($\sim 1^{\circ}$). Even for such a simple experiment, exposed to EAS from
a range of core locations and arrival directions, strong statistical 
differences are shown to exist between the pulse parameter
distributions of primary protons and those of heavier primary particles.
A range of primary energies can be investigated by varying the zenith
angle of observations. In this paper, results from simulations of
primaries in the energy range 20~TeV to 400~TeV are presented, although in
principle the technique could be extended to include the knee of the spectrum.
At the lower end of this energy range results can be compared to direct
measurements of the composition, while measurements at the upper end
can augment results from existing ground based experiments.

\end{abstract}
\end{frontmatter}

\newcommand{\C}{\v{C}erenkov }

\section{Introduction}

The chemical composition of cosmic rays measured at the Earth is an 
important key to understanding the production and propagation of 
cosmic rays. Up to $\sim$ 1TeV per particle the flux of cosmic
rays is sufficiently high that direct measurements of high
statistical significance can be made with satellite or balloon
based detectors. 
At energies of 100TeV per particle the flux of primaries is so low 
that direct composition
measurements are limited by large statistical uncertainties. 
Current knowledge of the composition at 100TeV, obtained from
direct measurement, is summarized in ~\cite{wat97}.

Above 100TeV primary fluxes are such that 
cosmic rays can only be studied through 
extensive air showers (EAS), 
generated as the primaries interact with the earth's atmosphere. 
At ground level EAS can be characterized by measuring secondary
electrons, muons, hadrons and {\C} light. If the primary
energy is sufficiently large ($>10^{17}$eV) fluorescence light
from nitrogen excitation is also detectable. 
On average, EAS from primaries of different mass will develop in
different ways, leading to composition dependent differences in the
secondary observables. In practice, inherent fluctuations in the
development of EAS and the complexity of interpreting ground
level measurements has limited the success of composition
measurement around the knee of the spectrum ($\sim 10^{15}$eV). 
Historically, the mass resolution of ground based experiments has been 
so poor that results are expressed as the ratio of light (protons and
helium) to heavy (mass $>$ helium) components. 
Estimates of this ratio at the knee from current experiments vary considerably
($\sim$0.3 to $\sim$0.6 ~\cite{wat97}) although there is general agreement 
that the average composition around the knee becomes heavier 
with increasing energy. Several new experiments, designed to
simultaneously measure many of the secondary observables of EAS, should
improve considerably the current knowledge of the cosmic ray composition
around the knee of the spectrum ~\cite{dic97,kla97,lin98,for97}.

\section{{\C} light from extensive air showers}
\label {eas}

The arrival time distribution of {\C} photons from EAS
has been  studied for a large range
of primary particle energies (see, for example \cite{hes99,chi99} and
references therein). For vertically incident primaries with energy 
$>$100TeV, which are detectable by ground level particle arrays, the 
vast majority of {\C} photons come from the electromagnetic (EM) 
component of the cascade. In this case the basic core-distance
 dependent time
structure of the {\C} pulse can be described by  the simple model
outlined in \cite{hil82}. Most of the {\C} emission occurs from
energetic particles traveling at speed $c$ near the core of the shower, 
which can be approximated as a single line of emission. The time structure is
determined by a combination of varying distances and refractive
index induced delays between the observer and different parts of the cascade. 
At the core, photons from the bottom of the shower will arrive first, with
photons emitted higher up being delayed by the refractive index of the 
atmosphere. Away from the core the {\C} photons emitted at the bottom of the
cascade experience greater geometrical delays than those emitted higher up. 
At the ``{\C} shoulder''(\cite{pat83})
 refractive and geometrical delays cancel and , in
this simple model, photons from all parts of the cascade arrive 
simultaneously. Beyond the {\C} shoulder the geometrical delays dominate
and the width of the pulse becomes a strong function of core distance.     
Clearly the greater the longitudinal extent of the shower, the
wider the {\C} pulse at most core locations. 

The simple model described above predicts reasonably well 
the general behavior of
{\C} pulses from EAS.
For real cascades, however, the relationship between core location and {\C}
pulse width is blurred by the distribution of particle
energies and the finite lateral extent of the shower core.
 The model also ignores
the contribution of {\C} light from muons. 

The highest energy muons are
created early in the hadronic core of the cascade and can easily
survive to produce {\C} light down to ground level. This light will
arrive in advance 
of light produced by the EM component of the cascade. 
The total muon energy of the cascade is carried by relatively 
few particles leading
to a poor efficiency in {\C} production compared to the EM
component. As the energy of the primary is reduced, however,
the relative contribution of the muons to the total {\C} yield is
increased. This is particularly true for the region inside the
shoulder of the lateral distribution, where many of the photons from the
the most deeply penetrating part of the EM  cascade arrive 
(\cite{pat83}). 
As the primary energy increases, the multiplicative nature of the EM
cascade efficiently converts the extra primary energy into large
numbers of {\C} producing electrons. The cascade   
develops deeper in the atmosphere, so the {\C} light is more concentrated 
at ground level and suffers less atmospheric absorption than light 
produced higher in the atmosphere.
While a higher energy primary also results in more energy in the muon
channel, much of that
energy is carried by a few very energetic muons or 
partly lost to the EM component if the charged pions 
interact rather than decay.

For a vertically incident primary hadron of a few TeV, the simple
model of {\C} pulse production described previously becomes 
inadequate. The electromagnetic component of the cascade will develop 
rapidly, and within $\sim$150m of the core the {\C} light
produced will appear as a ``flash'' 
inasmuch as the duration will be short compared to the duration
of the entire pulse.
The majority of the time structure of the pulse
comes from {\C} light from penetrating muons that appears on the leading
edge of the pulse. The ratio of light on the leading edge to that in the
``flash'' will reflect
the ratio of muons/electrons in the cascade capable of generating {\C}
light. 

The total time spread of {\C} photons observed within the
shoulder of the lateral distribution is
determined by the atmospheric thickness between the EM {\C}
emission and the observer. If observations are made at sufficiently
large zenith angles, the timing separation between EM and 
muonic {\C} light will be maintained for even a very energetic primary.

\section{The dependence of the pulse profile on the mass of the primary}

The effect of primary mass on the shape of the {\C}
pulse profile can be predicted through general arguments about 
EAS development: a detailed characterization of pulse profiles, obtained from 
Monte Carlo simulations, will be presented in section~\ref{montecarlo}.
Assuming maximal or near-maximal fragmentation of the primary nuclei, 
consider now
the differences in the development of the electromagnetic components
of EAS generated by protons and iron nuclei of the same total energy.
The longitudinal development profiles
of proton and iron induced EAS are remarkably similar (~\cite{lin98}). 
The individual
sub-showers from the nucleons of the iron primary develop and decay
more rapidly than the primary proton EAS, but these component nucleons
interact at a variety of atmospheric depths effectively
elongating the cascade. While the development of the
proton and iron cascades will have similar profiles, on average the
iron cascades will develop higher in the atmosphere. The transverse
momentum of the pions in a cascade increases only slowly with
total momentum (~\cite{wdo94}), so the lateral extent of the 
secondary particles in the iron
cascade will be greater than that of the proton cascade. The
combination of these two effects - height of maximum and wider lateral
distribution, result in the {\C} light from the EM component of the
iron induced EAS
being more diffuse at ground level than for the proton induced EAS. 
Over the energy range considered here, the {\C} photon density
at ground level for a primary iron nucleus is about half that of a primary 
proton of the same total energy.

The arguments used to describe the development of the EM cascade also apply
to some extent to the muonic component of the cascade: the muons in the iron
cascade tend to be produced higher and with greater lateral spread. The
muonic cascade from the iron primary is , however, much more efficient at
producing {\C} light. The energy of the muonic component of an iron induced 
cascade is carried by large numbers of relatively low energy muons. 
The much higher energy interactions at the
hadronic core of the proton cascade provides fewer muons with
larger average energy.
The overall result is that the ratio of the total {\C} light that 
is derived from the muons increases with increasing primary mass.

\section{Measuring {\C} pulse profiles} 

To fully exploit the mass dependent differences between EAS, the detector
must be able to collect enough photons 
to make a detailed  pulse profile for those EAS with 
EM components maximizing high in the atmosphere.
The bandwidth of the system must be high, and the field of view
sufficiently small that pulse parameterization is not 
seriously affected by
the night sky background. An isochronous large area mirror, such 
as those used in
TeV gamma-ray astronomy, viewed by a single photomultiplier tube would
fulfill these conditions. The use of such a system for cosmic
ray composition measurement has been described in \cite{rob98}, and
examined in detail for VHE cosmic rays (E$<$10TeV) in \cite{chi99}.

At any single zenith angle the range of primary energies that can be 
investigated is quite limited. The primary energy must be sufficiently
high that a large number of {\C} photons are available but the steep nature
of the primary energy spectrum and the shape of the {\C} lateral 
distribution bias any sample towards lower energy events. The 
higher the primary energy at a fixed zenith angle
the less distinct is the timing separation between the {\C} light of muonic 
and EM origin (see section ~\ref{eas}). A further consideration is
that for the higher energy events, the apparent image size is much larger 
so that on average less of the total angular distribution of the 
{\C} light is sampled by a narrow FOV
detector.

Fortunately the limited energy range is easily overcome by observing
at a range of zenith angles. The total atmospheric thickness changes
from $\sim$1000 gcm$^{-2}$ at zenith to $\sim$36000 gcm$^{-2}$ for
horizontal observations. This, in principal, would allow {\C}
composition measurements over a very large energy range (a few TeV to
tens of PeV). Observing at large zenith angles provides increased
collection area for the higher energy primaries, and also provides
a greater distance over which the {\C} emission can occur. This 
tends to stretch the pulse out, making the timing measurement easier
and less affected by systematic uncertainties in the measurement system.

A system similar to that described above has been operated on the BIGRAT
atmospheric {\C} detector. This system comprised a 4m diameter parabolic
mirror viewed by a single photomultiplier tube subtending a 
field of view (FOV)
of $\sim 1.0^{\circ}$. The system was designed to be sensitive to the 
differences between {\C} pulses initiated by gamma-rays and cosmic rays
for large zenith angle observations. While no 
detailed composition analysis
was performed, it was noted that the shape of the cosmic ray pulse
profiles was inconsistent with a pure proton composition (~\cite{rob98}).

\section{Monte Carlo Simulations}
\label{montecarlo}

The Monte Carlo simulations presented here 
have been made using CORSIKA version 4.5
~\cite{kna95}, with
GHEISHA code for low energy hadrons and VENUS for high energy hadrons.
The EM cascade is fully simulated using the EGS routines and Rayleigh, Mie
and ozone absorption processes are modeled for the {\C} light.
The detector consists of a single 5m diameter isochronous light collector
located at 160m above sea level. The mirror is viewed by a single
photomultiplier tube with assumed bialkali spectral sensitivity, 
subtending a full
FOV of $1.6^{\circ}$. This FOV has not been rigorously optimized for
pulse profile measurement: it is large enough that it can sample 
most of the angular distribution of the EAS of interest, and small enough
to exclude very large-arrival-angle large-core-distance cascades.
The photoelectrons detected by the 
photomultiplier are converted into a pulse by convolving the arrival time
of each photoelectron with a simple symmetric detector response function with
a rise-time (0-100\%) of 2ns. The waveform that is generated is sampled
4 times per nano-second. The night sky background is simulated by
adding Poisson distributed photoelectrons to the waveform at an
average rate of 2 per nano-second.

 \begin{table}
  \begin{center}
  \caption{Summary of the Monte Carlo simulation data set. The maximum
arrival direction for all primaries is limited to $2.0^{\circ}$ from
the center of the field of view.}
  \label{tab:mcsum}
  \begin{tabular}{cccc} \hline \hline
   primary  & zenith & minimum energy & maximum core distance \\
            &        &    (TeV)       &    (m)       \\ \hline
   proton   & $60^{\circ}$ & 15 & 450  \\
   helium   & $60^{\circ}$ & 20 & 450  \\
   oxygen   & $60^{\circ}$ & 20 & 450  \\
   iron   & $60^{\circ}$ & 30 & 450  \\
   proton   & $70^{\circ}$ & 100 & 720  \\
   iron   & $70^{\circ}$ & 200 & 720  \\
  \end{tabular}
  \end{center}
 \end{table}

\begin{figure}
\vspace{9cm}
\includegraphics{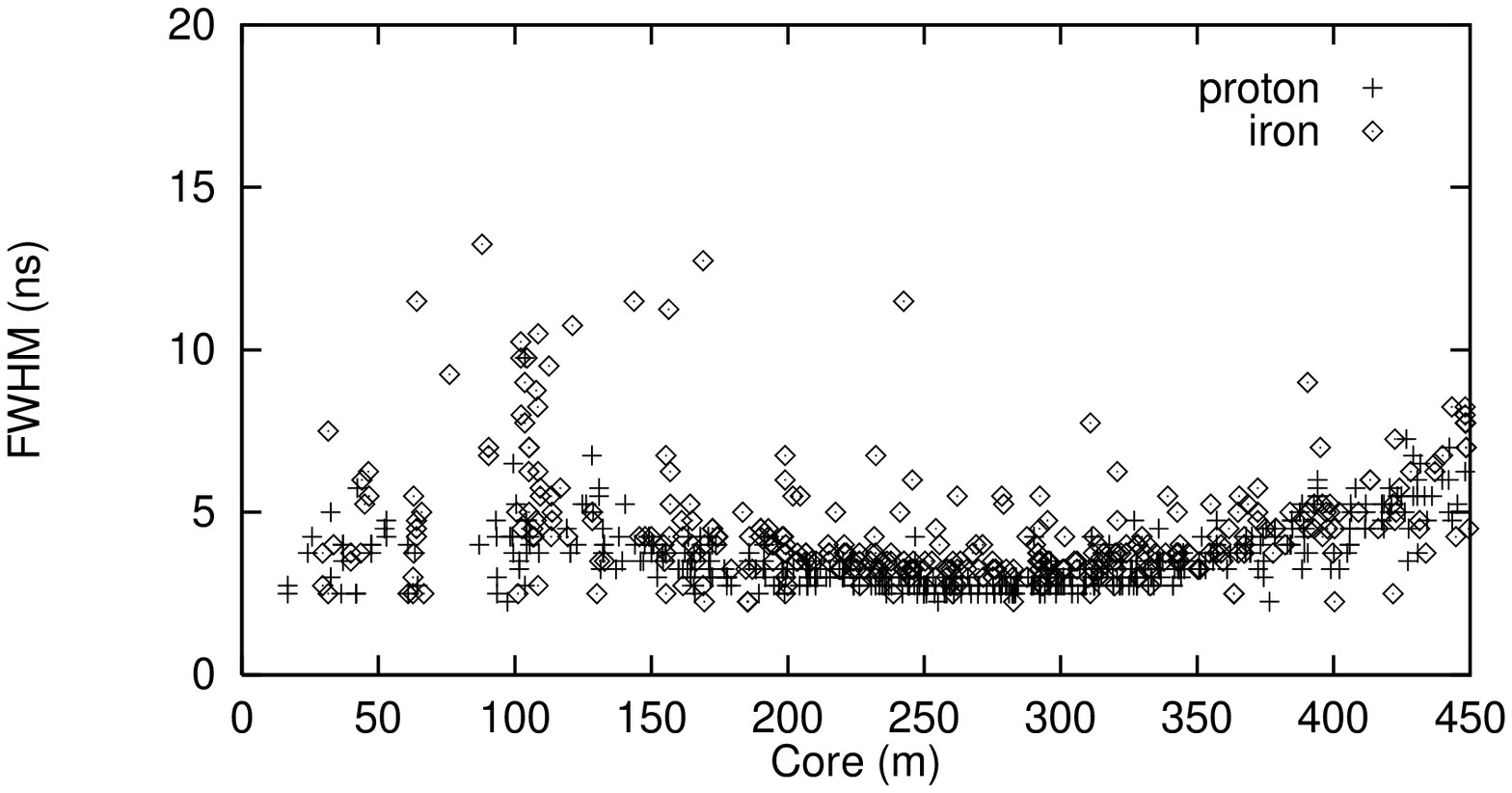}
\caption{FWHM of the {\C} light pulse versus core location 
for primary proton and
iron induced EAS at $60^{\circ}$ from zenith . The pulses contain between 600
and 900 photoelectrons (Monte Carlo simulation)}
\label{fwhm}

\vspace{9cm}
\includegraphics{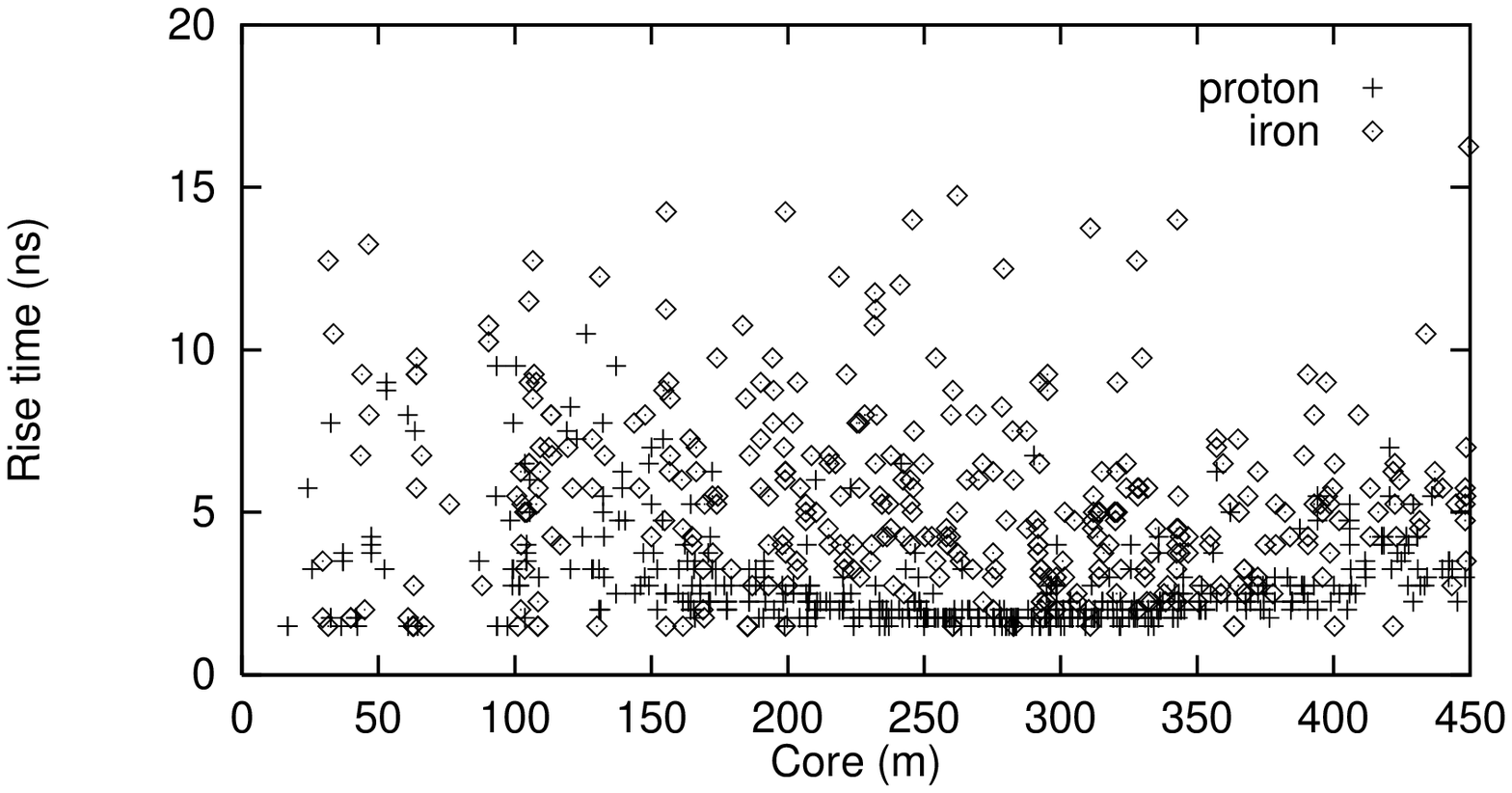}
\caption{Rise-time of the {\C} light pulse versus 
core location for primary proton and
iron induced EAS at $60^{\circ}$ from zenith . The pulses contain between 600
and 900 photoelectrons (Monte Carlo simulation)}
\label{risetime}

\end{figure}

In this paper, results of simulations at $60^{\circ}$ and
$70^{\circ}$ from zenith will be presented. At $60^{\circ}$ proton,
helium, oxygen and iron primaries have been simulated, but only
proton and iron at $70^{\circ}$ from zenith. To model a single
telescope realistically it is important to include primaries over
the full range of energies, core locations and arrival directions 
to which
the instrument is sensitive (see table~\ref{tab:mcsum} for a summary).
For all species an integral spectral index of -1.6 has been assumed.
To reduce computing time each shower has been sampled a total of
eight times.

At $60^{\circ}$ and $70^{\circ}$ from zenith the slant distances
are $\sim2$ and $\sim3$ vertical atmospheres respectively.  
It is possible to extend the energy range of observations up to the knee
region by observing at even larger zenith angles, but this is beyond
the limitations of the Monte Carlo simulation package used here. CORSIKA
v4.5 uses a flat earth/atmosphere and beyond $\sim 70^{\circ}$ this
leads to increasing inaccuracies in describing the depth profile
of the atmosphere. At extreme zenith angles the atmospheric depth
also changes considerably across the full angular acceptance of the detector
($\sim 4^{\circ}$), further complicating the interpretation of results.
As the total atmospheric depth traversed by the {\C} light increases, the
effects of atmospheric absorption become more important; this issue
will be addressed in more detail in section~\ref{experimental}.

\begin{figure}
\vspace{15cm}
\includegraphics{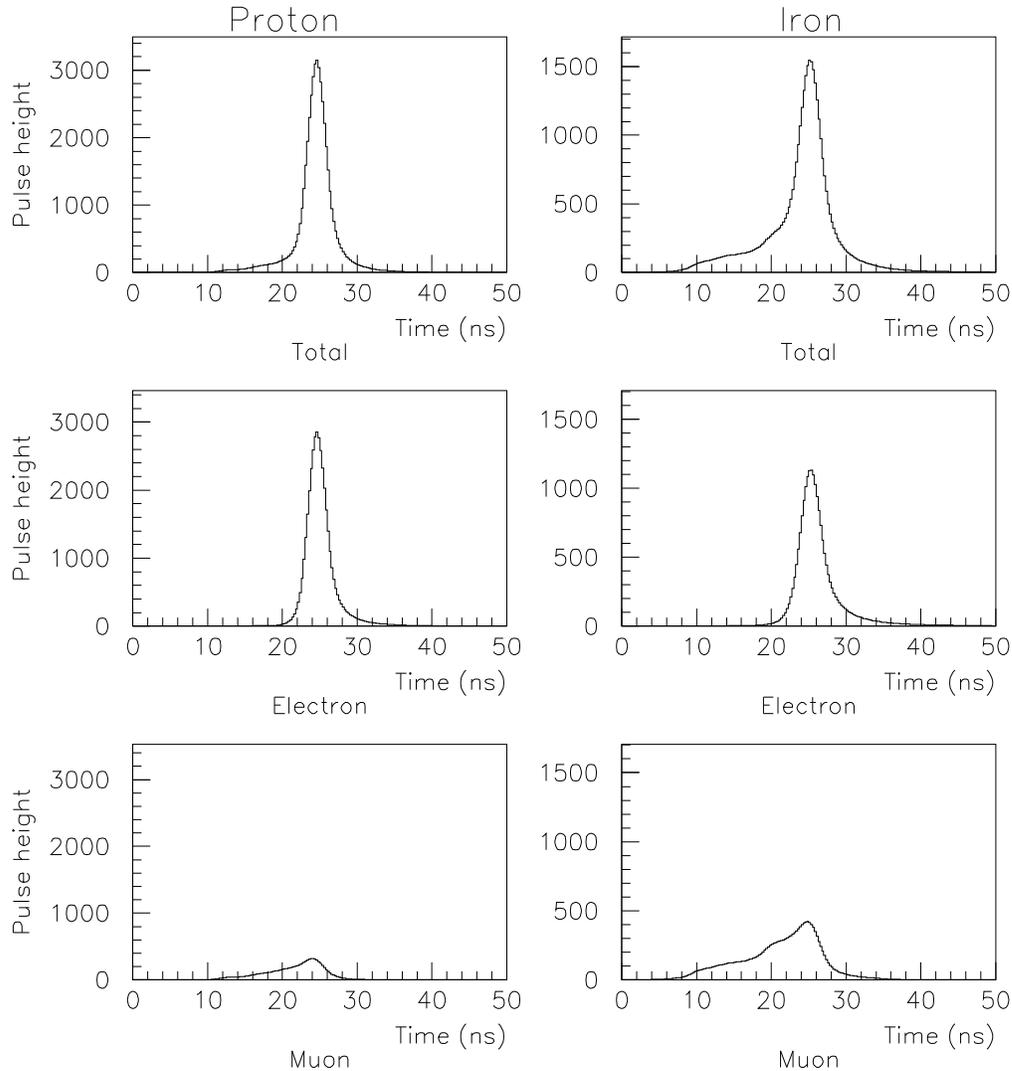}
\caption{Average {\C} pulse profiles for proton (left) and iron (right)
primaries. The individual pulses have been aligned to minimize the chi-square
differences between the {\C} light emitted by the electromagnetic 
component of each EAS. Also shown separately are the  contributions to
the average pulse profile from the muonic and electromagnetic components.
The units of pulse height are arbitrary.}
\label{pulses}
\end{figure}

Fig.~\ref{pulses} shows the average pulse profiles for proton and iron
primaries at $60^{\circ}$ from zenith. The pulses contain between
600 and 900 photoelectrons, but no other selection conditions have 
been applied. The pulse size selection acts to limit the range of
energies (and subsequently core locations) that are present in
the sample. 
The individual contributions to the {\C} pulse by
the muonic and EM components are also shown. It can be seen that
the muonic {\C} light is typically well in advance of the light
from the EM component and that the  muonic/EM {\C} light 
ratio of iron primaries is higher than that of proton primaries.

\begin{figure}
\vspace{18cm}
\includegraphics{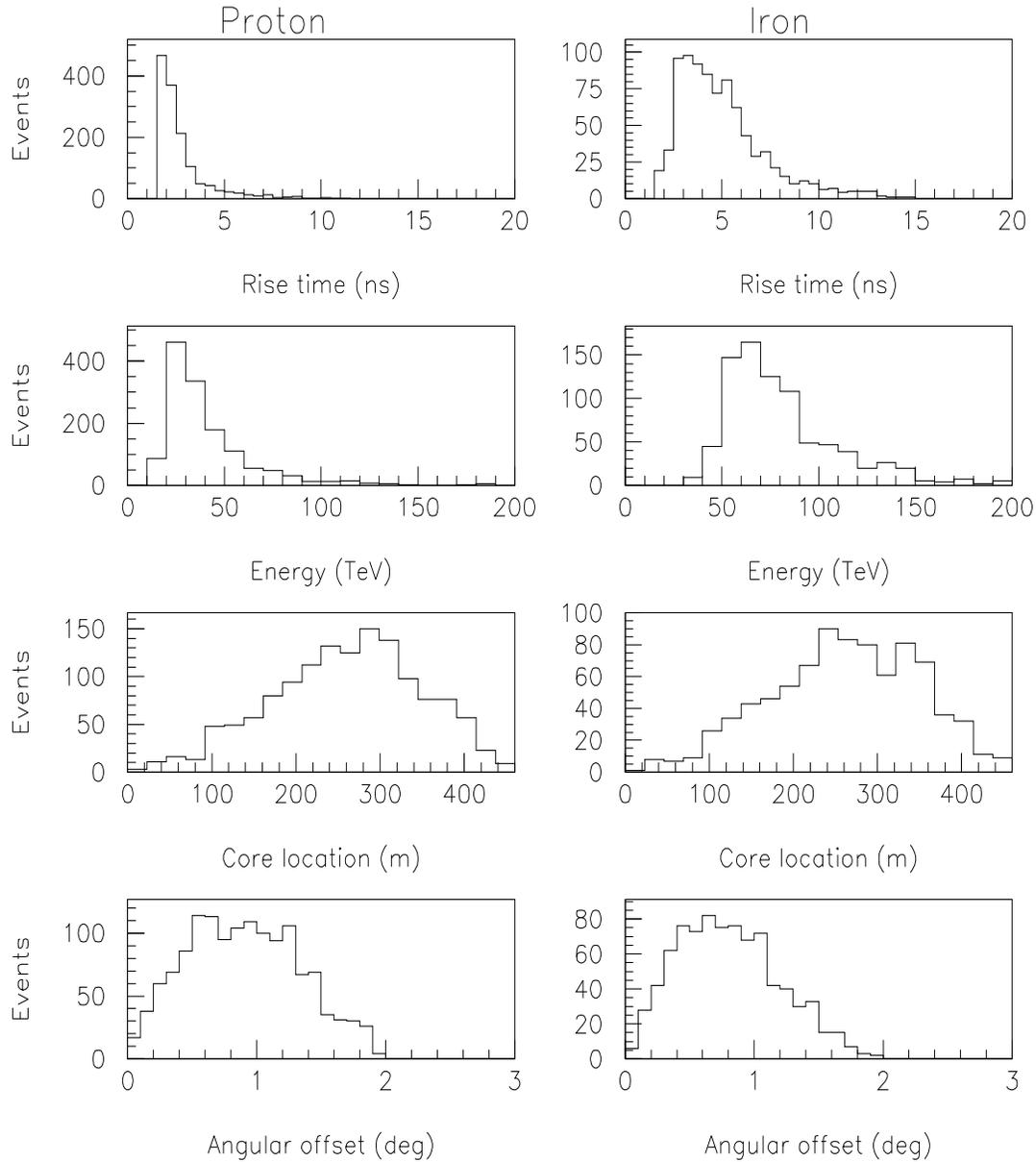}
\caption{Distributions of a number of different pulse and primary
parameters for proton and iron primaries at $60^{\circ}$ from zenith.
The ``Angular Offset'' describes the arrival direction of the primary
with respect to the center of the FOV.
The selection criteria, described in detail in the text,  applied 
to these samples are: 600 $<$ photoelectrons $<$ 900, FWHM $<$ 5.0ns, 
LT-ratio $>$ 0.1.} 
\label{params}
\end{figure}

\begin{figure}
\vspace{5cm}
\includegraphics{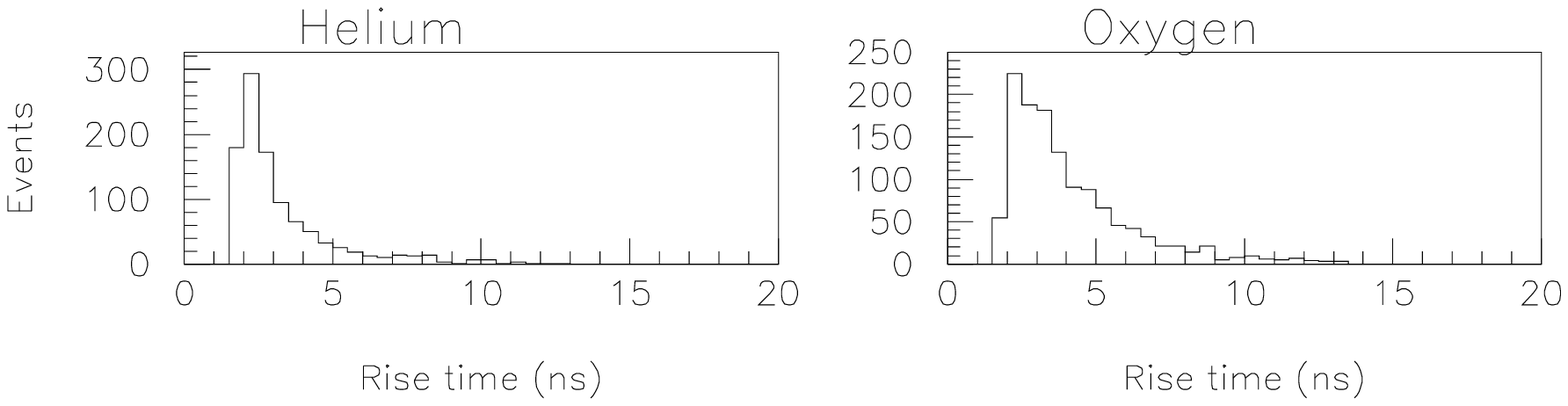}
\caption{The distribution of the {\C} pulse rise-time parameter for helium and 
oxygen primaries. The selection criteria used here are the  same as
those described in Fig.~\ref{params}.}
\label{he_and_ox_rt}

\vspace{15cm}
\includegraphics{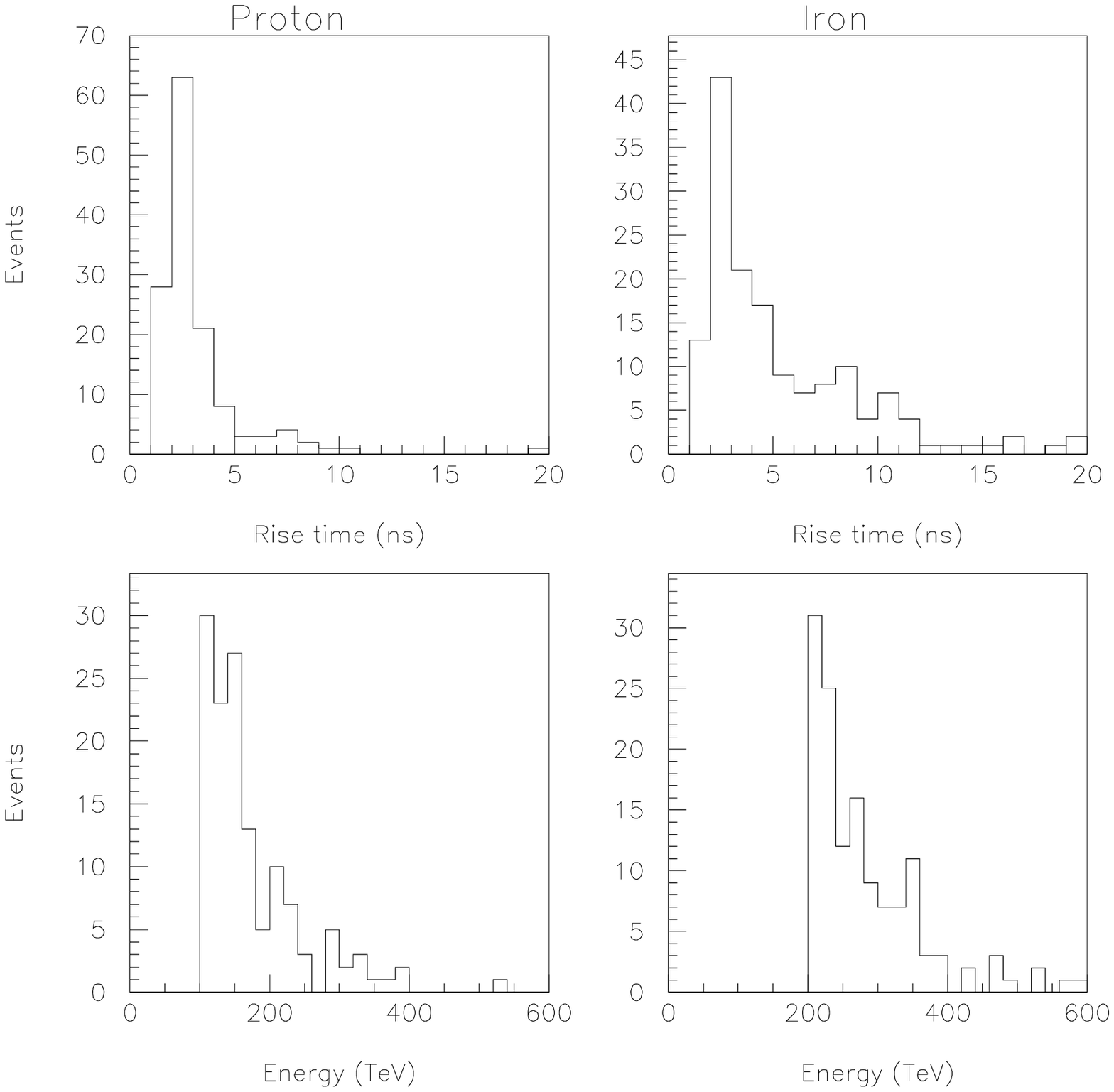}
\caption{The {\C} pulse rise-time and primary particle 
energy distributions of proton (left) and 
iron (right) induced EAS
at $70^{\circ}$ from zenith. Selection conditions: 
600 $<$ photoelectrons $<$ 900; FWHM $<$ 6.0ns, LT-ratio $>$ 0.1.}
\label{param70}
\end{figure}

The differences between iron and proton initiated {\C} pulse profiles 
can be seen in simple pulse parameters, such as rise-time (10\% to 90\%
of pulse maximum) and full width at half maximum (FWHM). The
distributions of these parameters as a function of core location are
shown in Fig.~\ref{fwhm} and Fig.~\ref{risetime}. In addition to 
rise-time and FWHM a third
parameter, called LT-ratio (Leading to Trailing signal ratio), 
will also be defined. The LT-ratio parameter is
the ratio of the signal on the leading edge of the pulse to the signal
on the trailing edge of the pulse. The signal on the leading and trailing
edges are calculated from the sum of photoelectrons arriving in
a 10ns period that starts 2.5ns and finishes  12.5ns 
from the maximum height of
the pulse. The LT-ratio parameter is useful for rejecting a small number of
events ($\sim$10\% of iron and $\sim$5\% of protons) where a large 
muon peak is present on the leading edge of the pulse.
This peak can cause a mis-characterization of the pulse by the 
simplistic determination of the rise-time
and FWHM parameters.

\begin{figure}
\vspace{14cm}
\includegraphics{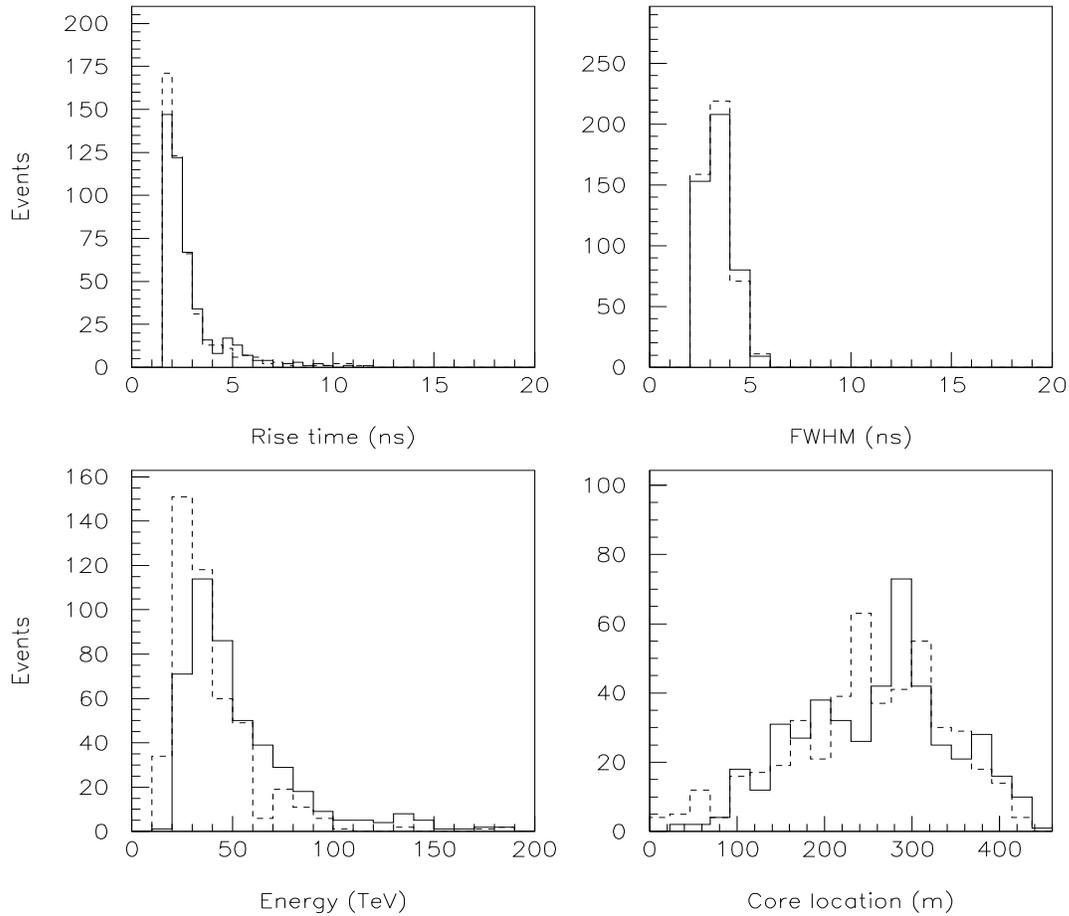}
\caption{The dependence, for protons primaries at $60^{\circ}$ from
zenith, of various 
pulse and primary parameters on
the depth of first interaction (DOFI). The solid line
is for proton initiated EAS with a DOFI less than 80 gcm$^{-2}$, the
dashed line for DOFI greater than 80 gcm$^{-2}$. Selection conditions
are those applied in Fig.~\ref{params}.}
\label{depth}
\end{figure}

The relationship between primary composition and rise-time is strongest
around the {\C} shoulder. The distribution of core locations can be limited
to some extent by making a simple FWHM cut (see Fig.~\ref{fwhm}). 
Fig.~\ref{params} shows the distribution of
rise-times and other parameters for proton and iron primaries after
pulses with FWHM greater than 5.0ns have been rejected. 
There are clear differences between the {\C}
pulse profiles of proton and
iron initiated EAS and this is reflected in the
 distributions of the rise-time parameter. Also
shown are the rise-time distributions of helium and oxygen primaries
at $60^{\circ}$ from zenith (Fig.~\ref{he_and_ox_rt}), and of 
proton and iron primaries at $70^{\circ}$ from zenith (Fig.~\ref{param70}). 

Many of the difficulties in interpreting EAS data at ground level are
due to the fluctuations in shower development. In particular, the
depth of first interaction (DOFI) variation for primary protons causes
large variations in the secondary particle properties at ground level.
Fig.~\ref{depth} shows that the {\C} pulse profile  
of a primary proton is largely independent of the DOFI.

\section{Composition estimation}

While clear differences exist between the pulse parameters of various
primary species, the interpretation of experimental results leading
to a composition estimate over a range of energies will clearly
be complex. Even for a narrow range of total pulse sizes at a
fixed zenith angle each primary species will have a different
distribution of energies, core locations and arrival directions.
As with other ground based experiments, correct interpretation of
results will rely on accurate modeling of cascade development, 
atmospheric attenuation and the detector response.

The considerable overlap between the rise-time distributions of the 
various primary species shows that it will be impossible to 
assign unambiguously primary mass on an event by event basis. Instead,
the composition may be inferred by combining the simulated rise-time 
distributions of individual primary species to reproduce the 
experimentally observed
rise-time distribution. The Monte Carlo simulations allow
the ratio of each species derived from such a comparison 
to be converted directly to a flux. 
If observations are taken over a range of zenith 
angles, such that the average energy at each zenith angle increases by a
factor of say, 5, the energy spectrum for each primary species 
can be inferred over
a wide range of energies. The assumed spectral index
for each species within each energy band can be adjusted by 
statistical resampling, and the comparison process repeated to
achieve consistency between the different energy bands.   


\begin{figure}
\vspace{7cm}
\includegraphics{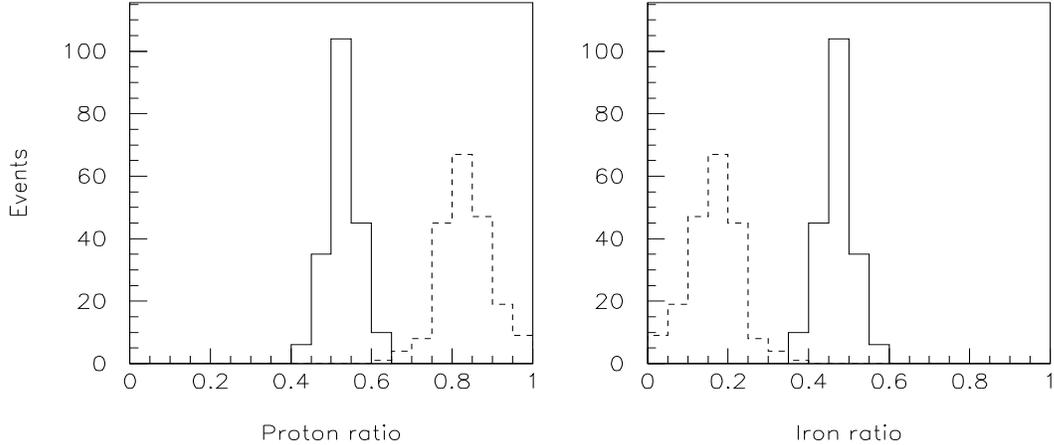}
\caption{The distribution in the predicted flux ratios of the 
primary species for actual flux ratios of (proton:iron=0.5:0.5) (solid lines)
and (proton:iron=0.8:0.2) (dashed lines)
(see the text for details).}
\label{final}
\end{figure}

An example of the accuracy to which the ratios of various
primary species can be estimated is shown in Fig.~\ref{final}.
This example, at  $60^{\circ}$ from zenith, represents the simplest case,
where the cosmic ray flux is assumed to consist of only protons
and iron nuclei.
The Monte Carlo
data set for each species has been divided randomly into two halves. From the
first half, a ``test distribution'' of rise-times has been created,
which will represent an experimentally measured sample. If the test
distribution is created assuming equal fluxes of proton
and iron primaries, after allowing
for triggering efficiency, collecting area and event selection, 
the ratios of events in the sample
are (proton:iron=0.79:0.21).   
The second half of the Monte Carlo rise-time data set
has then been repeatedly sampled, allowing the flux ratios of the primary
species to vary over
all possible values. Each of these ``sample distributions'' is then
compared to the ``test distribution'' using a Kolmogorov-Smirnov (K-S) test.
If the K-S test statistic indicates a 
probability greater than 90\% that the
test and sample distributions are drawn from the same parent distribution,
then the primary ratios are recorded. The most probable ratio for each 
species is determined with high precision, but the absolute accuracy
is limited - mainly by statistical fluctuations in the test distribution.
Fig.~\ref{final} shows the distribution of most likely ratios of
primary species 
for repeated regeneration of the test distribution. It should be noted that
each test and sample distribution is not fully independent, each being
drawn from a limited Monte Carlo data set. 
Each sampled 
distribution corresponds to only $\sim 10$ hours of
actual observations (400 events in each of the test and
sample distributions). A reasonable observational data-set of
several hundreds of hours duration, and more Monte Carlo simulations,
would provide greater flux accuracy than indicated in Fig.~\ref{final}.

The procedure described above can also be applied to a four
component cosmic ray flux (proton:helium:oxygen:iron), although
the flux accuracy is reduced compared to the two component
(proton and iron) fit. In addition, with the limited size
of the Monte Carlo data set, a completely unbiased search is
not possible, and the range of compositions searched must be limited
to avoid local statistical minimums in the differences between
the test and sample distributions.

\section{Experimental considerations}
\label{experimental}

One of the advantages of using a single mirror/single PMT
combination is the ease of calibration of such an experiment. The
mirror reflectivity, PMT quantum efficiency, gain and impulse
response can all be accurately determined. The background noise
to the {\C} pulses can be easily monitored and incorporated 
into Monte Carlo simulations. The
greatest source of uncertainty will be in characterizing the
atmosphere, and in particular describing the absorption of
the {\C} light in the atmosphere. Failure to correctly describe
the absorption profile of the atmosphere will distort the apparent
ratio of light emitted at varying depths from the observation point.

Demanding consistency of pulse parameter distributions on a night by
night basis should reject nights where the atmosphere is
disturbed (significantly different from a molecular atmosphere). 
In addition to this, atmospheric
attenuation could be measured directly through stellar extinction
and ground-level standard light sources placed at varying distances
from the observatory.
Although accurate accounting for absorption is most critical for 
observations at large zenith, the effects should also be observable for
near-zenith observations. It should be possible, therefore, to gauge
the accuracy of the absorption estimate and other calibration
procedures by comparing the {\C} pulse profile estimate of the
primary cosmic ray composition with that obtained by
direct measurement. This comparison should also be useful in
determining the accuracy of the Monte Carlo simulations as a whole.     



\section{Conclusion}

Monte Carlo simulations presented in this paper have shown that
the temporal distribution of {\C} light emitted from EAS is sensitive
to the muon/electron ratio of the cascade. Using a single large area
mirror coupled to a narrow field of view photo-detector, it is
possible to use these pulse profiles to estimate the chemical
composition of primary cosmic rays over a large range of energies.

\begin{ack}
The author would like to thank Philip Edwards, Jamie Holder, Bruce Dawson
John Patterson, Roger Clay and Gavin Rowell for helpful comments.
The author acknowledges the receipt of a JSPS postdoctoral fellowship.
\end{ack}

\end{document}